\documentclass{article}
\usepackage[utf8]{inputenc}
\usepackage{amsmath}
\usepackage{mathtools}
\usepackage{array}
\usepackage{xcolor}
\usepackage{graphicx}
\usepackage[a4paper, total={7in, 10in}]{geometry}
\usepackage{physics}
\usepackage{tabularx}
\usepackage{multicol}
\usepackage{hyperref}
\hypersetup{
	colorlinks=true,       
	linkcolor=blue,        
	citecolor=blue,        
	filecolor=magenta,     
	urlcolor=blue         
}
\usepackage{url}
\usepackage{caption}
\usepackage{subcaption}
\usepackage[sorting = none, backend = bibtex, style=numeric-comp]{biblatex}
\usepackage{tikz}
\usepackage{braket}
\usepackage{romannum}
\usepackage{authblk}
\usepackage{svg}
\usepackage{bm}
\usepackage{multicol}
\usepackage{graphicx,float}

\title{Influence of Orbital Angular Momentum of light on Random Spin-split modes in Disordered Anisotropic Optical media}
\date{}
\author[1]{Anwesha Panda}
\author[2]{Sneha Dey}
\author[3]{Yogishree Arabinda Panda}
\author[1]{Aditya Anurag Dash}
\author[4]{Aloke Jana}
\author[1,*]{Nirmalya Ghosh}
\affil[1]{Department of Physical Sciences\\
Indian Institute of Science Education and Research Kolkata\\
Mohanpur, India - 741246}
\affil[2]{Department of Physics, University of Calcutta, 92, APC Rd, Kolkata, India - 700009}
\affil[3]{Cluster of Excellence Physics of Life, Dresden University of Technology, Dresden, Germany - 01307}
\affil[4]{Department of Physics, Colorado School of Mines, 1500 Illinois St, Golden, CO 80401, USA}
\affil[*]{nghosh@iiserkol.ac.in}

\begin{document}

\maketitle

\begin{abstract}
\noindent
Spin orbit interaction of light in a disordered anisotropic medium is
known to yield spin split modes in the momentum domain because of the
random spatial gradient of the geometric phase of light. Here, we have studied the statistics of such spin split modes for beams carrying
intrinsic orbital angular momentum  through the quantification of
momentum domain entropy and investigated its dependence on various beam parameters. The influence of the spatial structure of the beam and the phase vortex on the statistics of the spin split modes were separately investigated using input Laguerre-Gaussian and Perfect Vortex beams passing through disordered anisotropic medium with
controlled input disorder parameter, which was realized by modulating
the pixels of a liquid crystal-based spatial light modulator. The
results of systematic investigations on  the impact of beam waist,
spot size and topological charge of the vortex beam shows that the
influence of the spot size on the emergence of the random spin split
modes is much more significant as compared to the other beam
parameters. 
\end{abstract}
\vspace{1 cm}
\begin{multicols}{2}
\section{Introduction} 
\vspace{0.1 in}
Spin-orbit coupling, also known as spin-orbit interaction (SOI), is an universal concept in physics that involves the coupling of spin and orbital degrees of freedom in both particles with mass (e.g., electrons) and massless particles (e.g., photons) due to relativistic effects. It occurs in various systems, ranging from atomic, condensed-matter systems to optical technologies, giving rise to interesting phenomena and potential applications. However, it is intriguing to note that this phenomenon, although typically discussed in the context of quantum particles, can also be manifested in classical light. This follows that classical light beams can carry both spin angular momentum (SAM) related to circular polarization and orbital angular momentum (OAM) associated with helical wavefronts of light, both of which interconvert into each other to create rich physics associated with SOI.
\vspace{0.6cm}
\par The SOI of light gives rise to two closely intertwined phenomena. The first phenomenon involves the influence of the light's trajectory on its state of polarization, resulting in the emergence of spin-dependent optical vortices which is commonly observed in systems with cylindrical or spherical symmetry. The second phenomenon is the reciprocal effect where polarization affects the trajectory of light. This effect is termed the Spin Hall effect (SHE) of light and is typically observed when symmetry is broken\cite{NaturePhotonics2015, shao2018spin, aiello2009transverse, ScienceApplications2015, Science2008, March2016, PhysRevA.75.053821, science.aap8640, Science2008,bliokh2008geometrodynamics}. SOI of light are basically of two types - one is geometric phase mediated and other is transverse angular momentum mediated. Geometric phases and their gradients, along with the preservation of the total angular momentum of light, are closely linked to optical Spin-Orbit Interaction (SOI) phenomena. Within this context, two distinct forms of geometric phase play a role—the spin redirection Berry phase and the Pancharatnam-Berry (PB) geometric phase. Spin Hall effect (SHE) originating from the geometric phase gradient, leads to either spatial domain or momentum domain shift. The other type of SHE, which is completely independent of the geometrical phase of light, originates from the transverse spin-angular momentum of light and is observed in the case of surface waves, evanescent waves, and waveguide modes, is not investigated in this paper
\cite{shao2018spin,aiello2009transverse,Xiao2017,ling2014realization,ScienceApplications2015,bliokh2015quantum,yin2013photonic}.
\vspace{0.6cm}
\par These advancements have resulted in a variety of fundamental effects in the realm of photonic SOI across diverse light-matter interactions. Remarkable phenomena like  spin-to-vortex transformation, orbital Hall effect, optical Rashba effect, plasmonic Aharonov–Bohm effect, spin-dependent transverse momentum, transverse SAM, spin-momentum locking, and spin-controlled directional waveguiding. These breakthroughs have paved the way for novel insights into universal SOI principles, offering new possibilities for designing spin-orbit photonic-devices.\cite{science.1234892,Ling_2017,Azimuthal,petersen2014chiral,aiello2015transverse,gorodetski2010plasmonic,yin2013photonic}
\vspace{0.6cm}
\par  Most of the scenarios described previously in the context of spin orbit interactions specifically those dealing with geometrical phases are for ordered inhomogeneous anisotropic medium. One of such realization is the metasurface which is fabricated by spatially structuring anisotropic media in the nanometer length scale. However, a recent discovery showcased the possibility of obtaining spin-orbit-coupled random scattering modes across the entire momentum range in a completely disordered, inhomogeneous, and anisotropic optical system. This phenomenon, known as the random optical Rashba effect\cite{Shitrit2013, Frischwasser2011}, is characterized by the presence of disordered spin-orbit coupling throughout the beam profile or a disordered spatial distribution of the geometric phase and its gradient \cite{science.aap8640}. Notably, the impact of intrinsic orbital angular momentum beams on spin-split modes remains unexplored, which is the focus of this investigation.
\vspace{0.6cm}
\par This paper aims to explore the influence of topological charge, spot size, and spatial structure of the beam on resulting spin-split modes in disordered anisotropic media. For this purpose, we have separately investigated using Lagaurre-Gaussian (LG) and Perfect vortex (PV) beams. The main purpose of using PV beams is to investigate the role of spot size and topological charges separately, as the size of vortices is independent of topological charges for PV beams. Here, we have quantified the statistics of spin-split modes in the momentum domain by the standard Shannon entropy. The study shows the momentum domain entropy of the spin-split modes are affected by both topological charge and spatial structure. 
\vspace{0.6cm}
\par The structure of this paper is as follows: Section 2 presents the theoretical framework of LG beams and perfect vortices, along with the formation of spin-split modes. Section 3 outlines the experimental procedure, while Section 4 discusses simulations and results of scattered modes in momentum space entropy. Finally, Section 5 provides concluding remarks and a summary of the findings.
\vspace{1.5cm}
\section{Theory}
\vspace{0.3 cm}
Let $\ket{E_i}$ and $\ket{E_o}$ be the input and output electric fields respectively, and the beam (containing RCP or LCP polarized light) is passing through an inhomogeneous(in the transverse plane with coordinates $\xi\rightarrow$x/y-plane and z being the propagation direction of light) anisotropic medium, we get --
\begin{equation}
    \ket{E_o} = e^{i{\phi_d(\xi)\pm\phi_g(\xi)}}\ket{E_i}
\end{equation}
where $\phi_d(\xi)$ is dynamic phases and $\phi_g(\xi)$ is PB geometric phase. 
When the two phases have equal gradient, $\frac{d\phi_d(\xi)}{d\xi}$=$\frac{d\phi_g(\xi)}{d\xi}$=$\Omega_{\xi}$, then the momentum domain shifts for RCP and LCP polarization states become 
\begin{equation}
    <{k_{\xi}}>_{RCP} = 2\Omega_{\xi}   \hspace{1em} and \hspace{1em} <{k_{\xi}}>_{LCP} = 0
\end{equation}where $$<k_{\xi=x/y}> = \frac{<E_0|i\frac{\partial}{\partial \xi=x/y}|E_0>}{<E_0|E_0>}$$ i.e., for RCP polarization, momentum domain shift becomes twice of spatial gradient of phase (geometric or dynamic), and for LCP, momentum domain shift does not occur \cite{ScientificReports2016,singh2020spin}. 
\vspace{1cm}
\par\textbf{Dynamical phase and Pancharatnam-Berry (PB) geometric phase in twisted nematic liquid crystal layers:} Polarized light gains PB geometric phase as well as dynamical phase while propagating in an anisotropic material. The dynamical phase for a linear birefringent medium is determined by the extraordinary and ordinary refractive indices ($n_e$ and $n_o$) and consequently it also depends upon the magnitude of linear retardance $\delta$ (defined as $\delta$ = $\frac{2\pi}{\lambda}(n_e - n_0 )d$, where $d$ is the path length and $\lambda$ is the wavelength). The PB geometric phase in such birefringent medium, on the other hand, is determined by the orientation angle of the anisotropy axis. Thus, in principle, one can produce equal spatial gradients of the dynamical
phase and PB geometric phase in an inhomogeneous birefringent medium by controllably and simultaneously changing the magnitude of linear retardance $\delta$ and the orientation angle of the anisotropy axis in the transverse plane. 
\vspace{0.6 cm}
\par The above can be realized by one of the readily available system, which is a twisted nematic liquid crystal-based spatial light modulator (SLM). The evolution of polarization in SLM can also be alternatively modelled using the effective Jones matrix ($J_{eff}$) as a sequential product of matrices of an equivalent linear retarder ($J_{reta}$, with effective linear retardance $\delta_{eff}$ and its orientation angle $\theta_{eff}$) and an effective optical rotator (with optical rotation $\psi_{eff}$). The rotation does not truly have a dynamical origin, it is actually related to the twist angle ($\psi$).\cite{ScientificReports2016,gupta2015wave,PhysRevLett.111.037802,DEV2012,Duran2005,Bahaa1990}
\begin{equation}
J_{eff} = R(\psi_{eff})J_{reta}(\delta_{eff},\theta_{eff})\end{equation}
\begin{center}
where $\psi_{eff} = -\psi + 2\theta_{eff}$
\end{center}

The total dynamical phase is primarily determined by the total linear retardance $\delta$, while the PB geometric phase is determined by effective optical rotation ($\psi_{eff}$). It was shown previously \cite{ScientificReports2016}, for a certain range of grey values (n=30-170), the two gradients, $\frac{d\delta(n)}{dn}$ and $\frac{d\psi_{eff(n)}}{dn}$ are equal. This leads to an equal gradient of geometric and dynamic phases which results in equation 2. The variation of the $\delta$, $\psi_{eff}$ parameters depends on the grey-level values (n). By changing the grey level values (n) from 30 to 170, both geometric and dynamical phases can be simultaneously tailored in an SLM. In our experiment we have modulated the pixels of SLM by generating random grey values (n) using a delta correlated uniformly distributed random function $f^{\epsilon}(\phi_g) = 1/2\pi\epsilon$ for $-\epsilon\pi \leq \phi_g(x) < \epsilon\pi$ otherwise $f^{\epsilon}(\phi_g) = 0$, $0\leq\epsilon\leq1, (where \epsilon$ controls the amount of randomness of $x$ coordinate).(shown in Fig.1)
\vspace{0.6 cm}
\begin{figure}[H]

\includegraphics[width=3.3in]{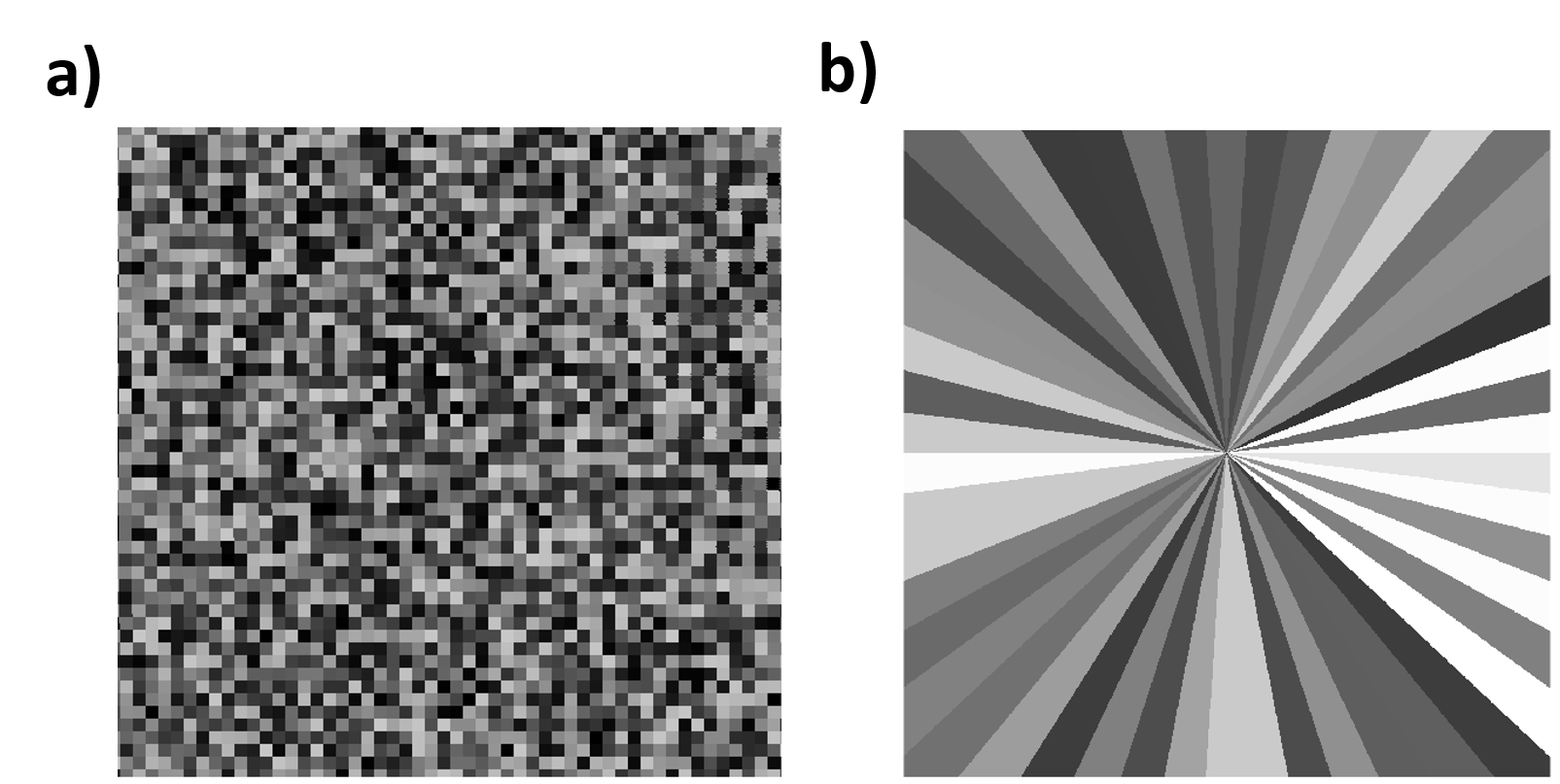}
\caption{\label{fig:1} The geometric phase distribution projected in the spatial light modulator (SLM). a) the grid was divided into pixels in x and y directions, where each pixel has a single random grey value (generated using delta correlated function as mentioned above). b) grid is divided azimuthally into n number of uniform divisions, with each azimuthal division having a single random grey scale value (n).
}
\end{figure}
\vspace{0.7cm}
\par The variation of phase in transverse plane, manifests as a distribution of intensities in momentum domain, which can be attributed to either dynamical phase, or geometrical phase, or a combination of both. Particularly, when this distribution is generated through the influence of only geometrical phase ($\phi = \pm \phi_g$), the SHE, becomes observable for right circularly polarized (RCP) and left circularly polarized (LCP) light. Here, the SOI of light results from the light beam's spatial inversion symmetry being broken by an inhomogeneous distribution of a geometric and dynamic phase combination\cite{NaturePhotonics2015, shao2018spin, aiello2009transverse, ScienceApplications2015, Science2008, March2016, PhysRevA.75.053821, science.aap8640, Science2008,ScienceApplications2015,loussert2013manipulating}. The strength of such an effect depends on the phase inhomogeneity acquired by the light beam. The corresponding intensity will be distributed throughout the momentum space as
\begin{equation}
    I (k_x,k_y)=|\iint^{+\infty}_{-\infty}e^{-i(k_xx+k_yy)}E_o(x,y)dxdy|^2.
\end{equation}
 \cite{singh2020spin}. 
It is also important to note, as long as we can describe the system(the anisotropic media) by a local phase gradient ($\xi$) , we will observe momentum domain spin-Hall effect of light as per equation 2. But as the randomness increases, beyond a certain value of disorderness when the system cannot be described by a particular local gradient anymore, one would observe random spin-split scattered modes in the momentum domain. To quantify the statistics of the momentum domain intensity distribution of the random spin-split scattered modes, we have defined the well-known Shannon entropy function in the following way-
\begin{equation}
    H = -\sum_i p_i[I(k_x,k_y)] log(p_i[I(k_x,k_y)])
\end{equation}
where $p_i[I(k_x,k_y)]$ is the probability density function (PDF) of intensity distribution of the scattered modes in the momentum domain\cite{science.aap8640}. 
\vspace{0.7cm}
\par For this purpose, in addition to the normal Gaussian beam we have used Laguerre-Gaussian (LG) and Perfect Vortex (PV) beams with different $l$ values (topological charges)  . Perfect vortex beams have a constant spatial structure regardless of their topological charge. This characteristic makes it a valuable tool for decoupling the effects of the spatial beam structure and the phase vortex.
\vspace{0.6 cm}
\par 
Laguerre-Gaussian beam at the source plane z=0 has the form -

\begin{equation}
    E_{pl}(x,y,0)=(\frac{\sqrt2\rho}{\omega})^lL_p^l(\frac{2\rho^2}{\omega^2})exp(-\frac{\rho^2}{\omega^2} + il\phi)
\end{equation}

where \(\rho = (x^2 + y^2)^{\frac{1}{2}}\) and \(\phi = tan^{-1}(y_0/x_0)\), $\omega$ is beam waist. $L_p^l$ is the associated Laguerre polynomial where p and $l$ are the radial modes and the angular modes (topological charges) respectively. 

\vspace{0.6 cm}
\par An approximate model of the PV beam \cite{Liu2017} to enable experiments is as follows:
\begin{equation}
    E_{PV}(\rho,\phi_0)=exp[-\frac{(\rho -\rho_0)^2}{\Delta\rho^2}]exp(il\phi_0)
\end{equation} 
\vspace{0.6 cm}
\par where ($\rho, \phi_0$) are the polar coordinates in beam cross section, $l$ is the topological charge $\rho_0$ is the radius of annular bright intensity, $\Delta\rho$ is a small width.
The Fourier transformation of an ideal Bessel beam function \cite{Vaity2015} may be used to calculate the approximate model of PV beams, and it can be written as
 \begin{equation}
     E_{BG}(\rho,z)= J_l(k_{\rho}\rho)exp(il\phi_0 + ik_zz)
 \end{equation}
 where $J_l$ is the first kind of $l^{th}$ order bessel function, $k = \sqrt{k_r^2 + k_z^2} = 2\pi/\lambda, r = (\rho,\phi_0)$ and $(k_r,k_z) $ are the radial and longitudinal wave vectors respectively.   
 \vspace{1.5 cm}
\section{Experimental Methods}
\vspace{0.3 cm}
A schematic of the experimental setup for observing spin-split scattering modes for LG beam is shown in Figure 2(a). A fundamental Gaussian mode of a 632.8 nm line of a He-Ne laser (HNL050L, Thorlabs, 5mW power) is used in this set-up. The beam is transmitted through the spatial light modulator (SLM, LC2012). A computer-generated fork hologram for different topological charges ($\pm$l) is projected to the SLM1 to produce an orbital angular momentum carrying beam (LG). After passing through the SLM1, the central beam (zeroth order) remains Gaussian, and its adjacent beams (first order) are LG with $\pm$l topological charges. An aperture is used after SLM to select and pass only one LG beam (+$l$ or -$l$) according to necessity. After being reflected by two mirrors (M1 and M2), the beam passes through PSG (Polarization State Generator) unit comprised of a Glan-Thomson linear polarizer (P1, GTH10M, Thorlabs, USA). The PSA (Polarization-State-Analyzer) unit consists of a linear polarizer and a quarter-waveplate, but positioned in the reverse order for selecting LCP or RCP polarization. In the middle of PSG and PSA, we have spatial light modulator (SLM2, same model as SLM1) realized as random inhomogeneous anisotropic media (shown in Figures 1a and 1b). Finally, there is a lens to obtain the Fourier image in the momentum domain. A CCD (1024$\times$768 square pixels, pixel dimension 4.65 $\mu$m, Thorlabs, USA) has been used at the end to collect the light at the Fourier plane. Spin-selective random scattering modes are observed for RCP polarization only and LCP polarization does not show any results. Figure 2(b) and 2(c) are the momentum domain intensity distribution for input LG beam while Figure 2(d) and 2(f) are for input PV beam . 
\vspace{0.6 cm}
\par In the usual case of spin-Hall effect of light, the input beam selected by PSG is linearly polarized. In SLM2, we have synchronously modulated the geometric and dynamic phases as shown in \cite{ScientificReports2016}. For one circularly polarized light (RCP), the geometric and dynamic phases add up; for the other, they cancel out giving no effect for disorder. This is called spin-selective scattering modes. To observe these modes, the inhomogeneous anisotropic medium was made possible by modulating the pixels of the SLM2 by randomly distributed grey values (n=30-170). The spin-selective scattering modes have been observed for all the beams such as Gaussian, LG, and perfect vortex. Figure 2(b),(c) shows this effect for LG beam with topological charge $l$=5 and Figure 2(d),(e) is for PV beam with topological charge $l$=5. \vspace{1cm} \par To build a setup for perfect vortex (PV), we need an additional lens and axicon in the above setup. Axicon lens converts LG to Bessel Gauss consequently Lens L1 has been used as a Fourier lens to transforms Bessel Gauss to perfect vortex. The position of lens L1 and axicon is commutative. PV beam is obtained exactly at the focal plane of lens L1 where the SLM2 [projecting the disordered anisotropic media (shown in Figure 1)] has to be placed accurately; a slight shift from this plane gives us bessel gauss. To obtain the Fourier transform lens L2 also has to be placed in the Fourier plane of L1. This was an experimental challenge to adjust L2 and SLM2 in the same plane, keeping the distance between them only less than 1 millimeter. We observed spin-selective scattering modes for perfect vortex as well similar to LG or Gaussian beam.
\vspace{0.6cm}
\par In order to calculate the momentum space entropy (Shannon entropy) in the Fourier plane of LG or PV beam, we calculate the normalized probability density function (PDF) of intensity distribution collected by CCD camera of 1024$\times$768 pixels. This normalized PDF is used in the expression of Shannon entropy using Eqn 5 \cite{science.aap8640}, to get the momentum space entropy. 
\end{multicols}

\begin{figure}[H]
\centering
\includegraphics[width=6.5in]{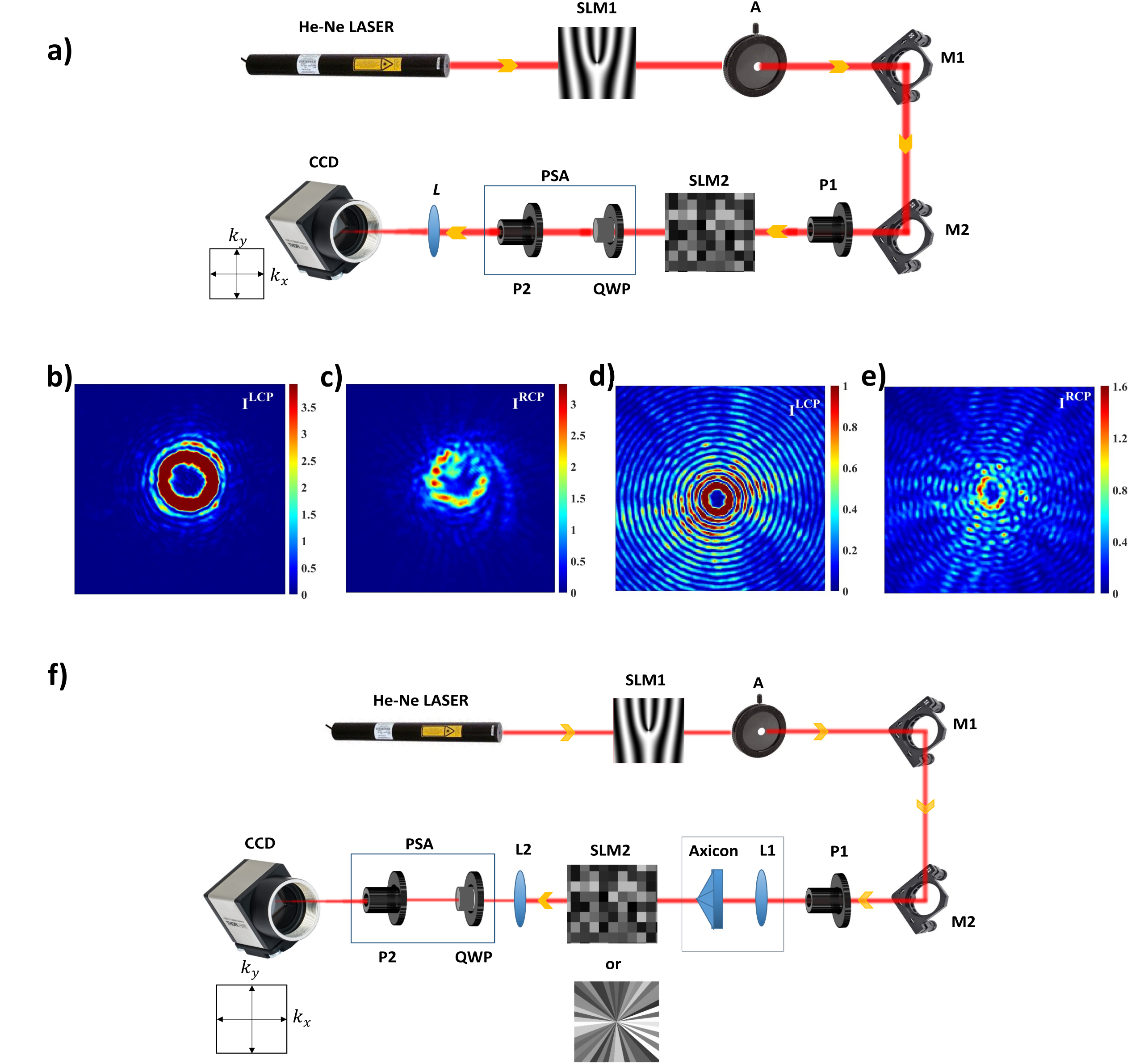}
\caption{\label{fig:2} \textbf{(a) Schematic of the experimental arrangement for observing spin-selective random scattering modes using Gaussian and Laguerre-Gaussian (LG) beams.} He-Ne Laser: light source; SLM1, SLM2: spatial light modulators (SLM1 was used to generate the LG beam with different topological charge, SLM2 was used to realize the disordered anisotropic media); A: aperture; M1, M2: mirrors; P1, P2: polarizers; QWP: quarter wave-plate; L: lens; CCD: camera (placed in the Fourier plane to record momentum domain) here. The polarization-state analyzer (PSA) unit comprises a linear polarizer and quarter waveplate. This unit analyzes desirable polarization states of light (LCP or RCP in this case). Spin-selective scattering modes of LG beam transmitted through disordered anisotropic media of $\epsilon$=1 are shown for topological charge $l$=5 for RCP \textbf{(b)} and LCP \textbf{(c)} by projecting 50 azimuthal divisions in SLM2 (each division has different grey-scale values (n) which have been assigned randomly). Similar result are shown for input PV beam of topological charge $l$ = 5 passing through the same grey scale pattern projected on SLM2 as mentioned above; in \textbf{(c)} LCP and \textbf{(d)} RCP. The spin selective property of the random inhomogeneous media \cite{singh2020spin} holds not only for Gaussian beam but also for LG and PV beams. \textbf{(f) Schematic of the experimental setup to observe spin-selective scattering modes of input perfect vortex beams} Most components are similar to the setup shown in (a).  Additional components to produce the PV beam are:- L1, Axicon. L1 , L2: lenses which are used as Fourier lenses; Axicon lens converts LG beam to Bessel Gauss.
}
\end{figure}
\vspace{1.5cm}

\begin{multicols}{2}
\section{Results and Discussions}
\vspace{0.3 cm}
At first, we consider the effect of input LG beam with the varying $l$ values and input disorder parameter $\epsilon$ of the disordered anisotropic optical media. As shown in Figure 3, which displays the experimentally observed momentum space intensity distributions of Gaussian and LG beam with $l$=0, 3, 5, and with varying input disorder parameters, $\epsilon$ = 0, 0.5, 1. It is observed as the input disorder parameter $\epsilon$ increases, the number of random scattered spin-split modes increases gradually which indicates increase in momentum space entropy. This variation has been quantified by using equation 5 and is presented in Fig 4.

\begin{figure}[H]

\includegraphics[width=3.5in]{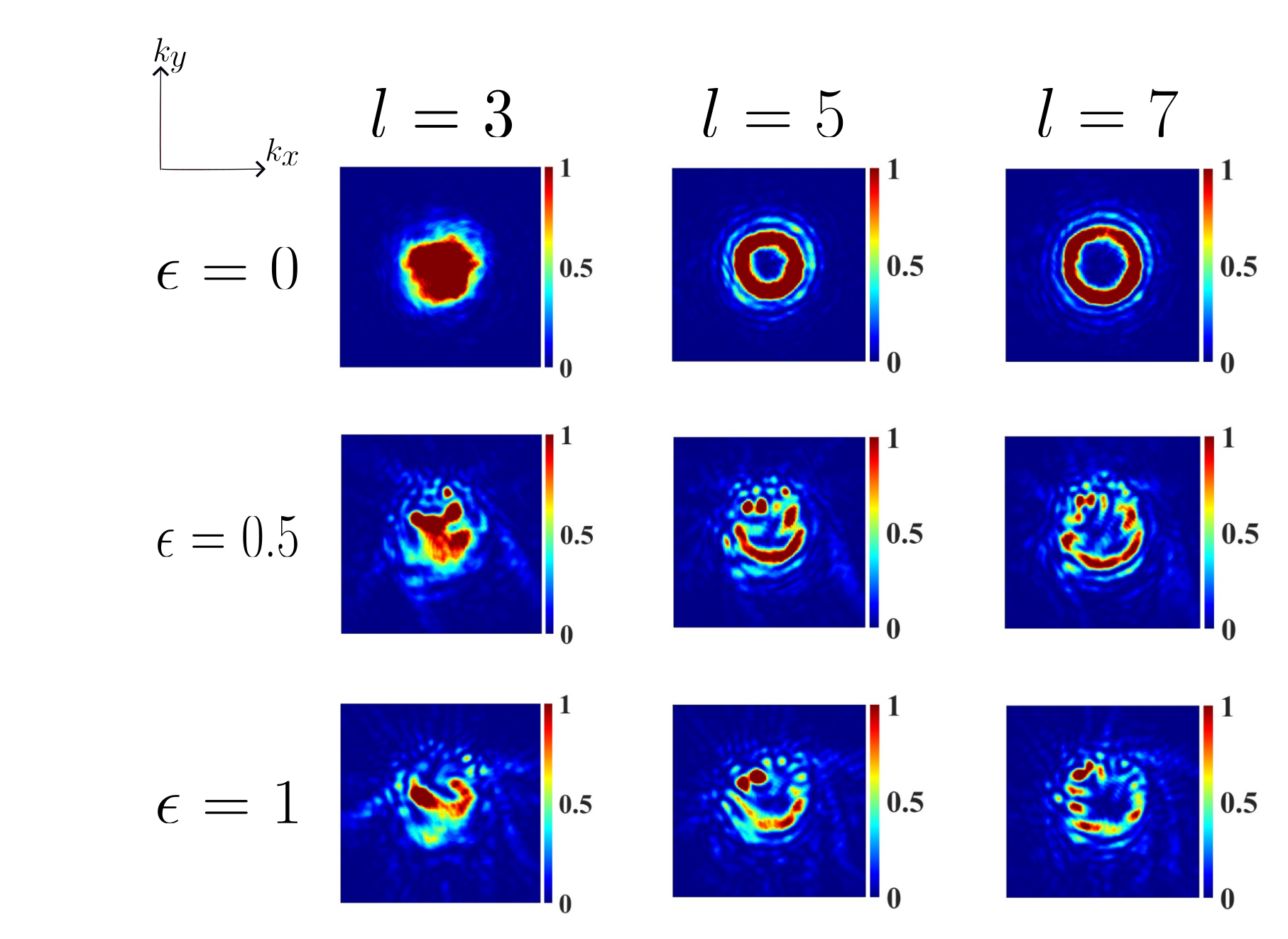}
\caption{\label{fig:3} Variation of momentum domain intensity distribution of Gaussian and Laguerre-Gaussian beams for different topological charges  ($l$=0, 3, 5) transmitting through disordered anisotropic media  with input disorder parameter $\epsilon$ = 0, 0.5, 1. As the $\epsilon$ value increases, the beam gets more and more scattered. SLM2 was projected with 50 azimuthal divisions of random grey scale (n). Results shown are only for RCP polarization selected using PSA. 
}
\end{figure}

\par Figure 4, depicts the variation of momentum space entropy with input $\epsilon$ parameter along with varying topological charges $l$ for input LG beam. We observe a sudden surge in momentum space entropy at a critical value of the disorder parameter $\epsilon$ i.e., there exists a threshold value of $\epsilon$. This provides strong evidence for existence of a phase transition within the system under study. This occurs because, as long as a local spatial phase gradient (both geometrical and dynamical) can be clearly defined within the system, we anticipate the Spin Hall Shift. However, as the heterogeneity of the optical media intensifies, a critical point is reached where the local spatial phase gradient can no longer be distinctly defined. i.e., the media becomes completely random. We observe the emergence of random spin split modes. In Fig. 4 (a) and (b) we see a clear threshold value of the disorder parameter $\epsilon$ around 0.25. After this, as input disorder parameter $\epsilon$ increases the variation of momentum space entropy increases rapidly. Also, we see greater $l$ values imply greater total momentum space entropy for a input disorder parameter $\epsilon$. It is vital to note that as one varies the $l$ values, one would not only vary the topological charge but also the effective vortex size or the input spot size of the beam in the spatial domain. This correspondingly changes the amount of inhomogeneity probed by the input beam in the spatial domain and consequently their momentum domain span. It may so happen that $l$ influences the spin split modes but in addition to that the effective spot size will also affect them as larger spot size will probe greater area of inhomogenity.  This is why we anticipated that not only $l$ but the spin split modes will also be affected by the spot size of the input beam. Thus, it is necessary to investigate the role of beam spot size separately.   
\begin{figure}[H]
\includegraphics[width=3.5in, height = 4 in]{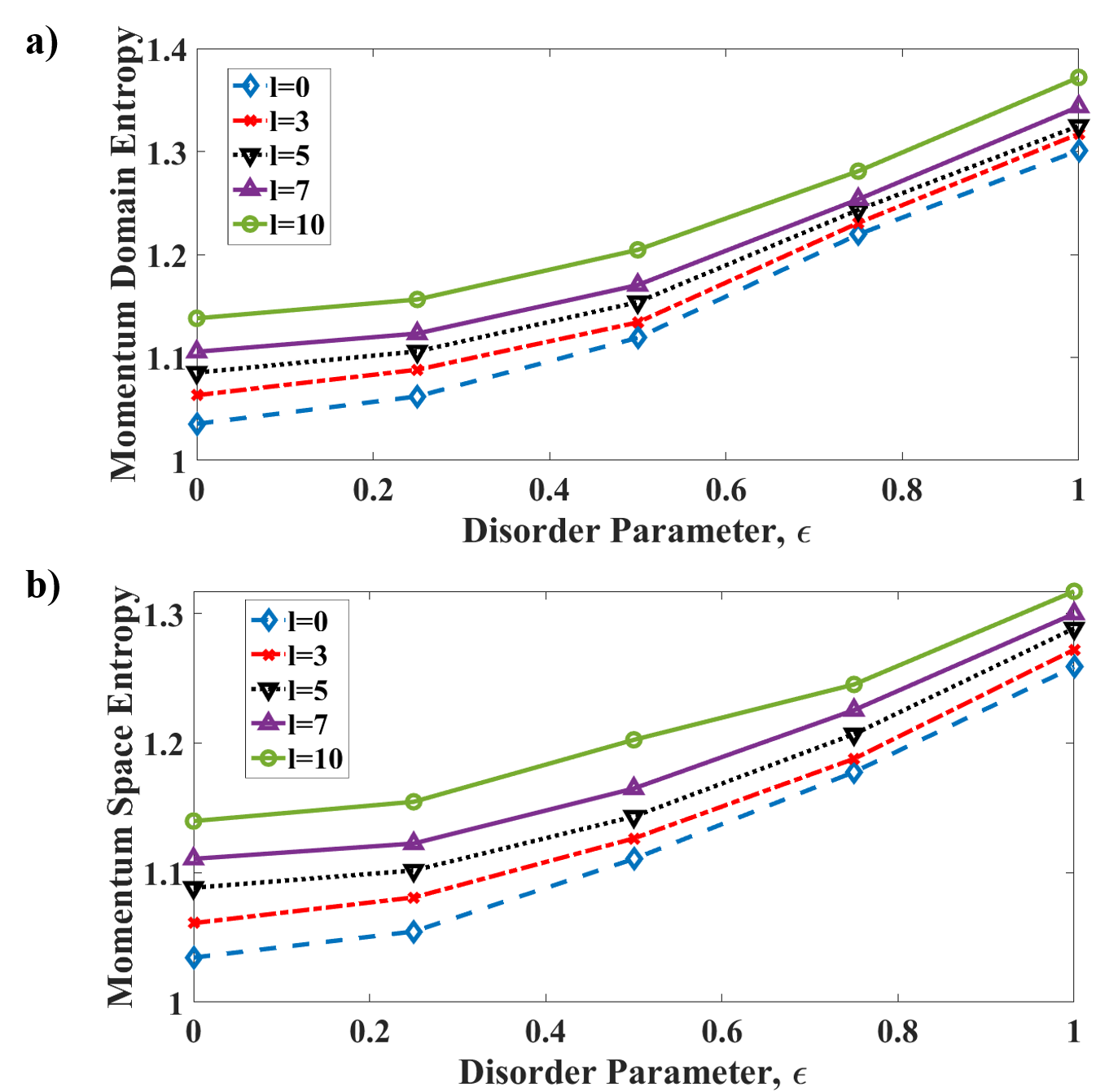}
\caption{\label{fig:3} Momentum space entropy ($H$) variation of LG beams with different topological charges , $l$ = 0, 3, 5, 7, 10 , passing through disordered anisotropic media while disorder parameter $\epsilon$ varies from 0 to 1 . The results are for different effective beam waist in the spatial domain, (a) 2.5mm  and (b) 1.88mm . The LG beam is passed through the SLM2 which is projecting with Figure 1(a) random media. Results are shown only for RCP polarization, was selecting using PSA.
}
\end{figure}
\par We have investigated the role of spot size of the input beam by keeping the $l$ value same and varying the effective spot size of the input beam. The momentum space entropy variation shown in Fig 5. follows the same trend as in Fig 4. A clear threshold is observed here for disordered parameter around $\epsilon$ = 0.25, similar to the previous case. However, the variation of momentum space entropy with the input disorder parameter $\epsilon$, is much steeper when we have larger spotsize of the input beam. 
The important concept here is that if we take the same heterogeneous system and use a smaller spot size of the beam then it probes a lesser inhomogeneity but the corresponding momentum span is large. On the other hand, for a larger spot size the scale of inhomogeneity probed by the input beam is more, for a corresponding small momentum domain span. So, one would expect a higher momentum space entropy in the latter case. Thus, the effects of spotsize and topological charge $l$ on the spin split modes are still combined.
\vspace{0.3 in}
\begin{figure}[H]
\includegraphics[width=3.5in, height = 3.6 in]{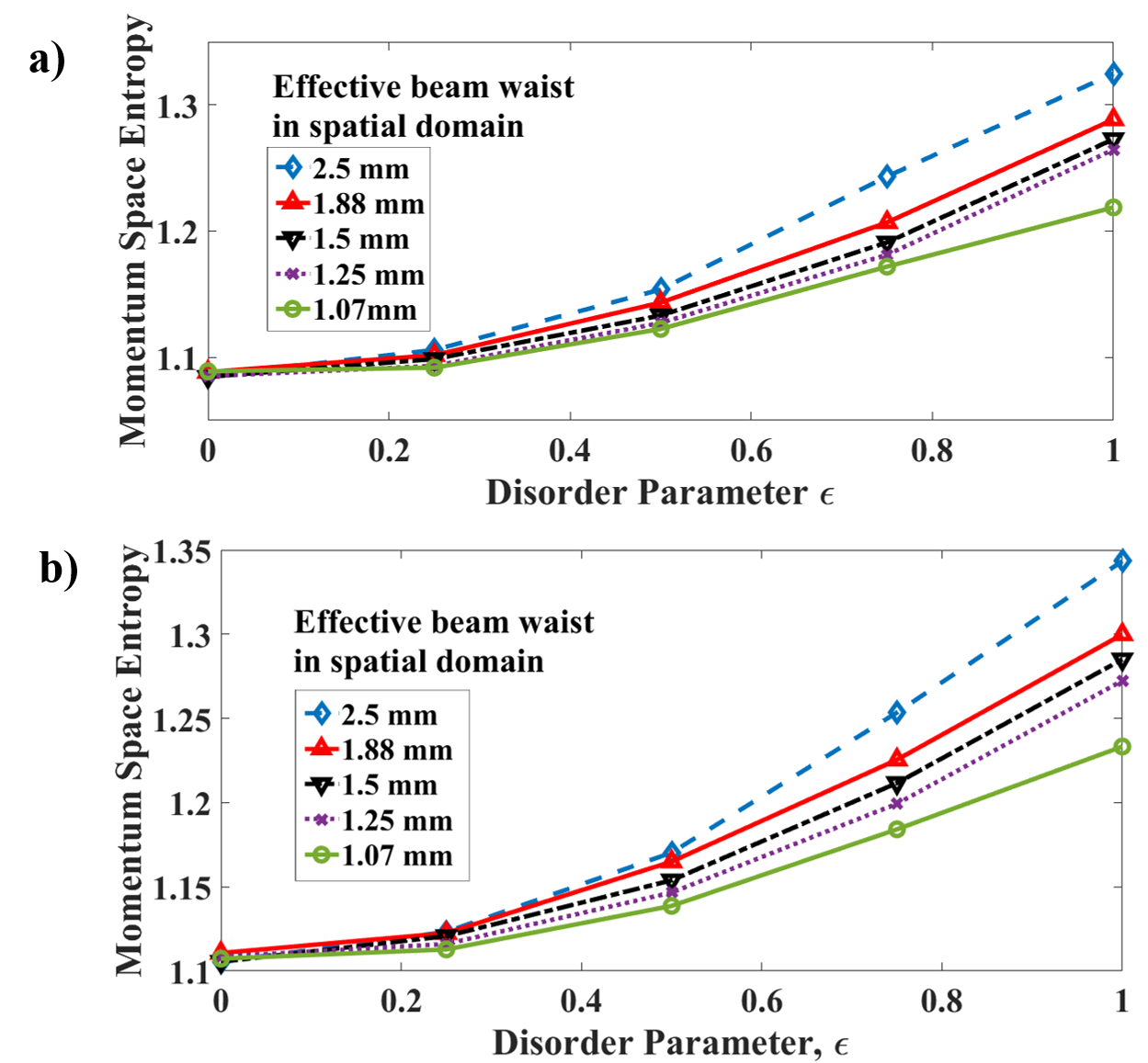}
\caption{\label{fig:4} Momentum space entropy ($H$) variation of LG beams for different effective beam waist in spatial domain, $\omega_0$=2.5mm, 1.88mm, 1.5mm, 1.25mm, 1.07mm; for (a) $l$=5  and for (b) $l$=7  for  disorder parameter $\epsilon$ varying from 0 to 1. The LG beam is passed through the SLM2 which is projecting with Figure 1(a) random media. Results are shown only for RCP polarization, was selecting using PSA.}
\end{figure}
\vspace{0.3 in}
\par To decouple their effects, we have used the Perfect Vortex (PV) beams. As mentioned in the theory section the spatial structure of the PV beam remains constant for all $l$ values. The experimentaly obtained PV beams are shown in figure Fig. 6 (a). The figure shows for topological charges $l$ = 3, 5, 7 the spatial structure of the beam remains constant in spatial domain. The corresponding momentum space intensity distribution for $l$ = 3, 5, 7 for input disorder parameter $\epsilon$ = 0, 0.5, 1 is shown in Fig. 6 (b). Similar to the previous case, with increasing input disorder parameter $\epsilon$, the beam gets more scattered. For the maximum input disorder parameter, for $\epsilon$ = 1,  the maximum intensity of the central beam decreases as the number of random spin split modes is highest for this case. Since spatial structure for PV beams is independent of $l$, the inhomogeneity probed by the input PV beam remains fixed here for all values of $l$; unlike the case of LG beam, where the inhomogeneity probed by the input beam changes with the change of $l$ values due to change in the spatial structure of the beam. 
\vspace{0.3 in}

 \begin{figure}[H]
\includegraphics[width=3.5in, height = 3.7in]{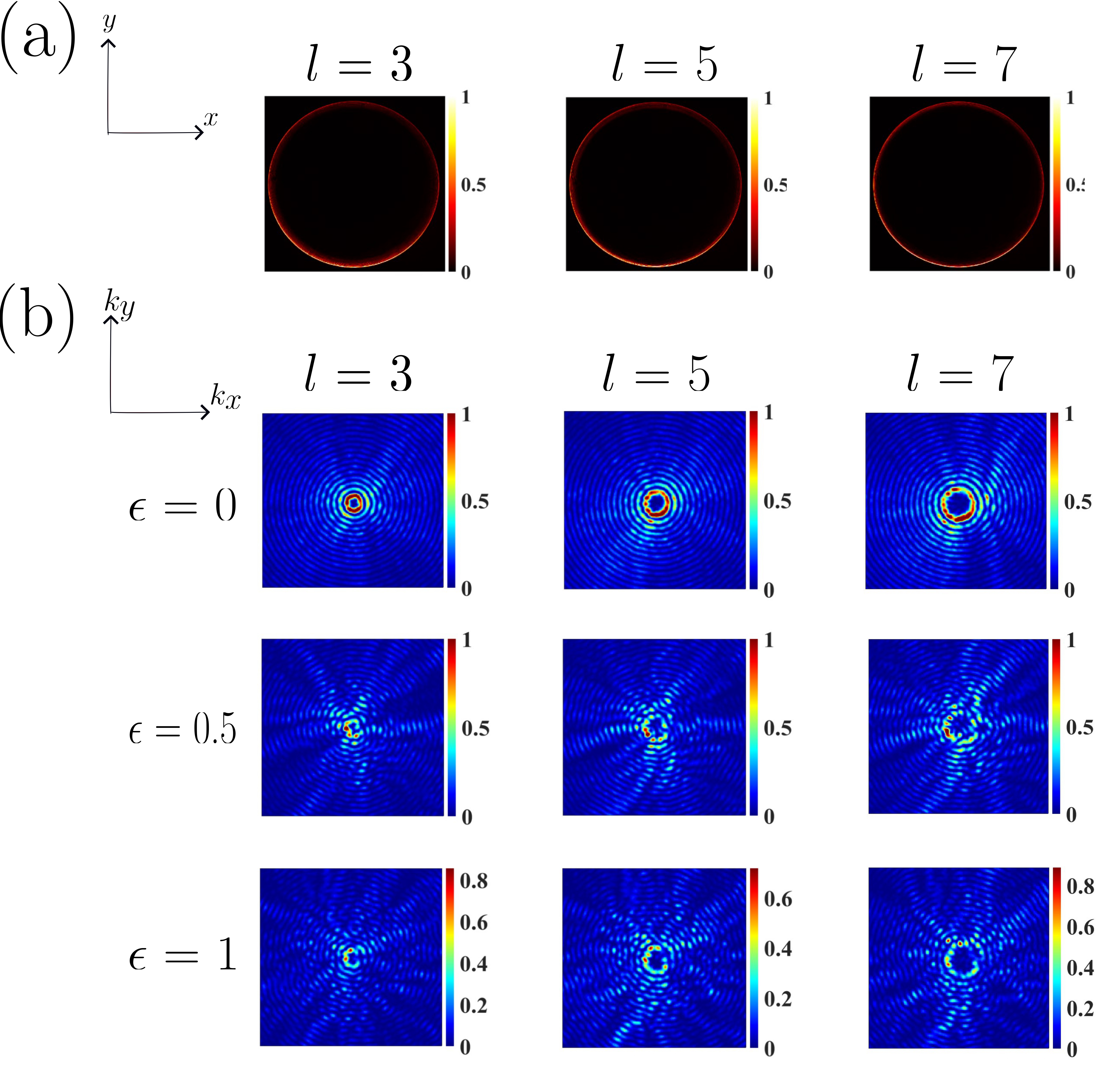}
\caption{\label{fig:5}(a) Input Perfect Vortex beam in spatial domain where the spatial structure of the beam remains same for different topological charges $l$=0, 5, 10. (b) The corresponding momentum space intensity distribution after the input perfect vortex beam with the $l$ = 3, 5, 7 passes through inhomogeneous anisotropic media (projected in SLM 2) with disorder parameter $\epsilon$ values = 0, 0.5, 1.}
\end{figure}
The corresponding momentum space entropies are quantified in figure 7. Even though in the space domain, the amount of heterogeneity probed by the input PV beam remains same for all $l$ values, their momentum space (Fourier domain span) may get changed as PV transforms into Bessel-Gauss beam in the Fourier domain. Despite this what we observed is with variation in values of $l$ and with variation in disorder parameter $\epsilon$, the change of momentum space entropy is rather minimal.
Therefore, this indicates, that the influence of topological charge $l$ alone on the statistics of momentum domain spin split modes and on their corresponding momentum domain entropy is rather weak. For the LG beam, the momentum domain entropy of the spin split modes is also primarily
dominated by the change in spatial structure of the beam i.e. the effective spot size rather than its topological charge. This is further confirmed by the simulation in Fig 8.
\begin{figure}[H]
\includegraphics[width=3.7in,height = 2.2 in]{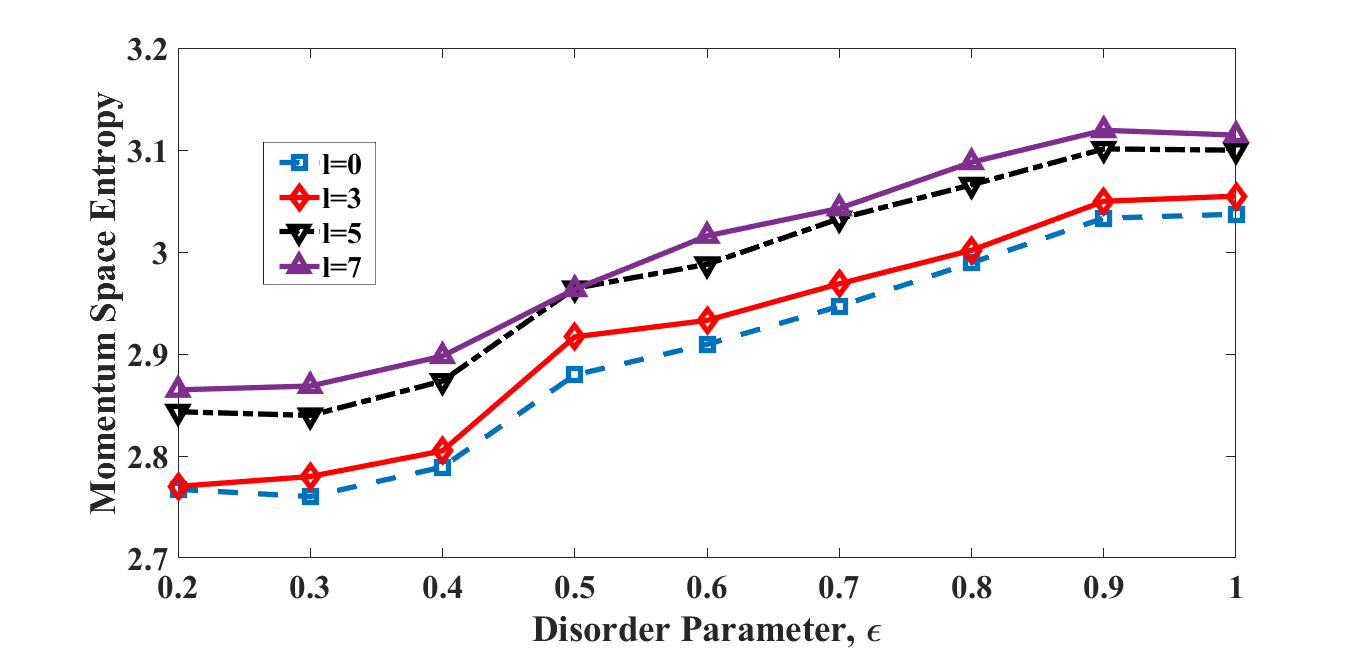}
\caption{\label{fig:7}The variation in momentum space entropy as we change the input disorder parameter ($\epsilon$, ranging from 0.2 to 1) using PV beams with distinct topological charges ($l$ values of 0, 3, 5, 7) as input beams. The optical medium consists of 300 azimuthal divisions, each division is assigned a random grey value generated by the delta correlated function as mentioned before, and these results are for right-circularly polarized (RCP) light.
}
\end{figure}
Fig. 8 shows the variation of momentum space entropy with the variation input disorder parameter for LG beam of different input spot sizes. Fig 8 (a) corresponds to input beam waist 10 mm and Fig. 8 (b) corresponds to input beam waist of 2.5 mm. The momentum space entropy is much higher for smaller input spot size: 2.5 mm (fig. 8 (b)) than for the larger input spot size: 10mm (fig. 8 (a)). It is known that effective spot size of LG beam decreases with lowering in $l$ value. Here, we clearly see a rise in threshold value as the overall input spot size increases. As discussed before, a smaller input spot size probes smaller area of inhomogeneity but corresponds to a larger momentum domain span. Thus, we can conclude from fig. 8 that the role of input spatial structure of the beam is more dominant than its topological charge ($l$).

\begin{figure}[H]
\includegraphics[width=3.5in]{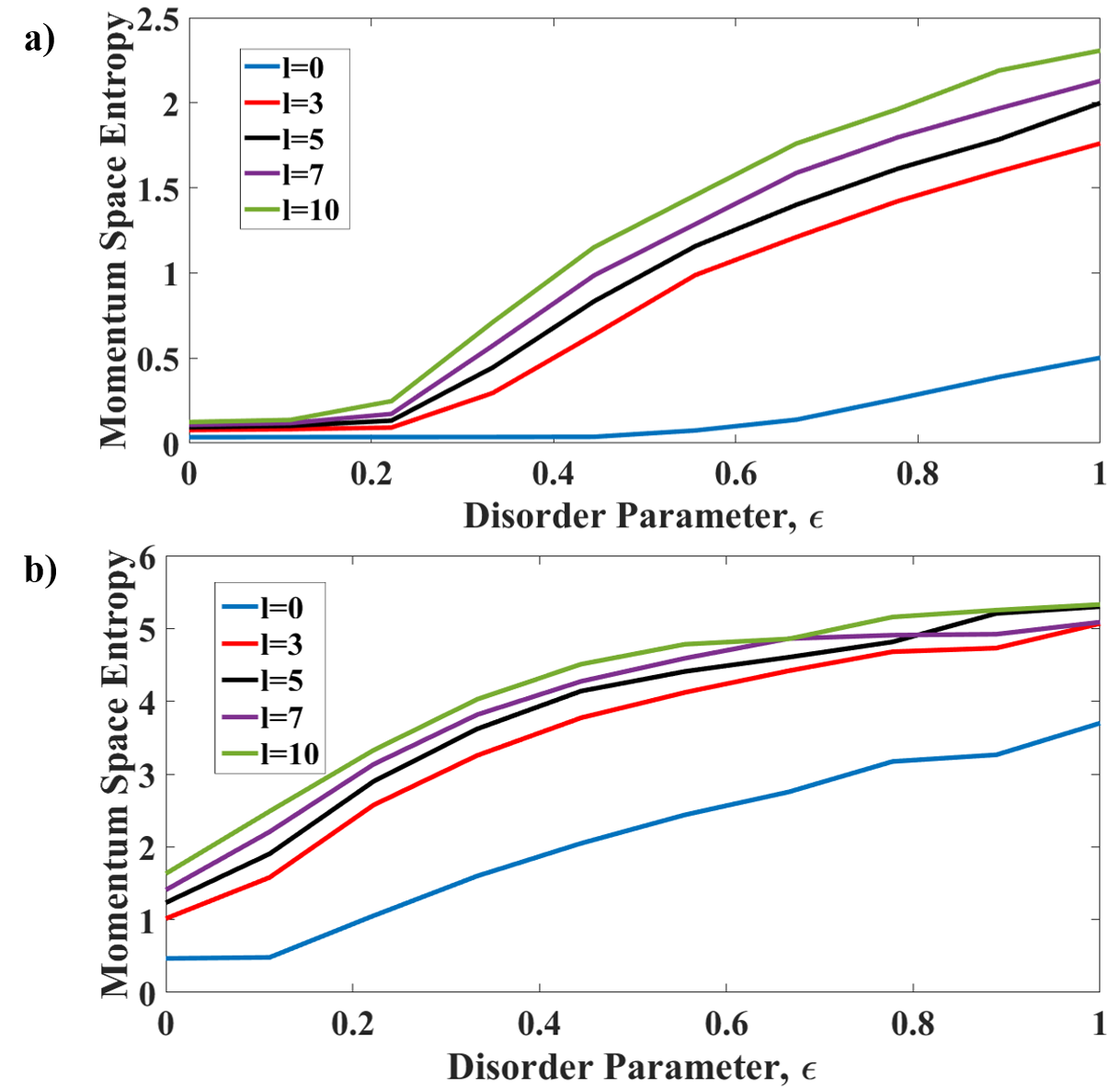}
\caption{\label{fig:8}Simulation showing momentum space entropy variation of the randomly scattered spin split modes when LG beams (of different topological charges) of different beam waists pass through the disordered inhomogeneous anisotropic optical media. (a) beam waist = 10mm in the spatial domain. (b) beam waist = 2.5 mm (smaller spot size compared to the previous one but corresponding to a larger Fourier domain span). As the overall input spot size increases, the threshold value increases gradually. 
}
\end{figure}

\section{Conclusion}
We have investigated the role of topological charges ($l$) and the beam waist ($\omega_0$) in orbital angular carrying beam on the momentum space entropy of disordered system. Our investigation shows when we are dealing with both LG and perfect vortex beam, the spot size plays an important role in momentum space entropy $H$ of spin-split modes compared to topological charge $l$.
\end{multicols}

\printbibliography
\end{document}